\documentclass[apj,a4paper,12pt,useAMS]{emulateapj} %===== HERE for astro-ph (2nd choice)
\usepackage{amsmath} %==================================== HERE for astro-ph (2nd choice)

\usepackage{graphicx,color}
\usepackage{natbib}

\newcommand{\simgt}{\lower.5ex\hbox{$\; \buildrel > \over \sim \;$}}
\newcommand{\simlt}{\lower.5ex\hbox{$\; \buildrel < \over \sim \;$}}

\def\singlebond{\@makechembond\@ne}
\def\doublebond{\@makechembond\tw@}
\def\triplebond{\@makechembond\thr@@}

\newcommand{\vect}[1]{\boldsymbol{#1}}
\newcommand{\matr}[1]{\mathbf{#1}}

\shortauthors{Liao et al.}

\shorttitle{Polarization Calibration with GBT}

\begin{document}

\title{
Accurate Polarization Calibration at 800 MHz with the Green Bank Telescope}

\author{
Yu-Wei~Liao\altaffilmark{1}, Tzu-Ching~Chang\altaffilmark{1}, Cheng-Yu Kuo\altaffilmark{1,2}, Kiyoshi Wesley Masui\altaffilmark{3,4}, Niels Oppermann\altaffilmark{5}, Ue-Li Pen\altaffilmark{5}, and Jeffrey B. Peterson\altaffilmark{6}
}

\altaffiltext{1}{Institute of Astronomy and Astrophysics, Academia Sinica,
11F of Astro-Math Building, AS/NTU, 1 Roosevelt Rd Sec. 4, Taipei 10617, Taiwan; ywliao@asiaa.sinica.edu.tw}
\altaffiltext{2}{Department of Physics, National Sun Yat-Sen University, 70 Lienhai Road, Kaohsiung 80424, Taiwan}
\altaffiltext{3}{Department of Physics and Astronomy, University of British Columbia, Vancouver, British Columbia, V6T 1Z1, Canada}
\altaffiltext{4}{Canadian Institute for Advanced Research, CIFAR Program in Cosmology and Gravity, Toronto, Ontario, M5G 1Z8, Canada}
\altaffiltext{5}{Canadian Institute for Theoretical Astrophysics, University of Toronto, 60 St. George Street, Toronto ON, M5S 3H8, Canada}
\altaffiltext{6}{McWilliams Center for Cosmology, Carnegie Mellon University, Department
of Physics, 5000 Forbes Ave., Pittsburgh PA 15213, USA}

\begin{abstract}
Polarization leakage of foreground synchrotron emission is a critical
issue in HI intensity mapping experiments.  While the sought-after HI
emission is unpolarized, polarized foregrounds such as Galactic and
extragalactic synchrotron radiation, if coupled with instrumental 
impurity, can mimic or overwhelm the HI signals.  In this paper we present
the methodology for polarization calibration at 700-900 MHz, applied on
data obtained from the Green Bank Telescope (GBT). We use
astrophysical sources, both polarized and unpolarized sources including quasars and pulsars, as calibrators to
characterize the polarization leakage and control systematic effects in
our GBT HI intensity mapping project. The resulting fractional errors on 
polarization measurements on boresight are well controlled to within $0.6\%$-$0.8\%$
of their total intensity. The polarized beam patterns are measured by
performing spider scans across both polarized quasars and pulsars. A dominant Stokes $I$ to $V$
leakage feature and secondary features of Stokes $I$ to $Q$ and $I$ to $U$
leakages in the 700-900 MHz frequency range are identified. These characterizations are important for separating foreground polarization leakage from the HI 21 cm signal.   

\end{abstract}


%=======================================================================================================1
%=======================================================================================================1
%=======================================================================================================1
\section{Introduction}\label{sec:intro}

Neutral hydrogen (HI) is one of the most promising probes of the 
high-redshift universe. It can be used to uniquely trace the matter
distribution at early times well into the dark ages and the cosmic dawn
era, reveal the cosmic reionization process and shed light on the
complex astrophysics in early galaxy formation, and probe the large-scale
structure at late times, allowing measurements of the geometry and
structure growth of the universe. To obtain statistical measurement 
of the three-dimensional structure of HI in emission, the intensity mapping 
technique has been advocated (e.g., \cite{Chang2008},
\cite{Loeb2008},  \cite{Chang2010}). Making use of the redshifted 21 cm HI
intensity mapping dataset obtained at the Green Bank Telescope (GBT), \cite{Masui2013} and
\cite{Switzer2013} have measured the HI cross-power spectrum with the WiggleZ optical galaxies and
reported limits on the HI auto-power spectrum, respectively, in the frequency range of
700-900 MHz or a redshifted HI range of $0.6 < z < 1$. Combining the cross-power and auto-power spectrum,
the neutral hydrogen fluctuation amplitude, $\Omega_{\mathrm{HI}}b_{\mathrm{HI}}$ has been constrained as $\Omega_{\mathrm{HI}}b_{\mathrm{HI}}=[0.62^{+0.23}_{-0.15}]\times 10^{-3}$.

The main challenges of HI intensity mapping at these redshifts include synchrotron
foreground radiation and man-made radio frequency interference (RFI). Localized
synchrotron emission from extragalactic sources and diffuse
synchrotron emission from the interstellar medium (ISM) in the Milky Way
are the dominant astronomical foreground signals in the frequency range of
interest. In the GBT HI observing fields reported by \cite{Switzer2013}, which are at high Galactic
latitudes, the synchrotron emission is still three orders of magnitude
brighter than the 21 cm signals. However, synchrotron radiation is expected
to be spectrally smooth (\cite{Oh2003}; \cite{Seo2010}). If all the
instrumental effects, including calibration, spectral response, and
primary beam pattern, are well understood and controlled, the
synchrotron foregrounds will then have fewer degrees of freedom than the HI
signals along the line of sight, or along the frequency direction, and
can be separated. In the GBT data, RFI is found to be removable or
controllable by flagging of frequency channels.

A crucial instrumental effect that needs to be controlled is
polarization leakage. Although HI emission is considered 
unpolarized, the leakage of polarized synchrotron foreground emission
into total intensity via imperfect instrumental response could
introduce excess power and extra degrees of freedom into the observed
intensity signal. Simulations suggest polarized intensity of Galactic Foregrounds
can contain frequency structure via Faraday rotation induced by the
ISM;  leakages of such spectrally fluctuating polarization intensity into total
intensity could thus mimic the HI signals along the frequency or
redshift direction (\cite{Jelic2010}, \cite{Moore2013}). Furthermore, polarization
leakage itself may not be a smooth function of frequency.  Therefore,
one may not be able to simply separate polarization leakage from HI signals by isolating spectrally smooth patterns.

The polarized Galactic foreground is one to two orders of
magnitude fainter than total intensity in regions with low Galactic
synchrotron emission at the frequency of interest. The amplitude in
total intensity due to polarization leakage would be comparable to
that of the HI signals if the leakage fraction is at the percentage level, which is common in single dish radio telescopes \citep{Marti2012}. Therefore, polarization calibration needs to be performed very carefully to eliminate the contamination from polarized foreground signals.

Recent investigation of LOFAR polarization leakage by
\citet{LOFAR2015} provides 
a good estimate of polarization calibration errors (less than $0.005\%$) of inflicted on the HI signal of interest at 150 MHz.
As an interferometer, the redundancy of LOFAR baselines dramatically
reduces the integrated polarization calibration errors; this is however 
not the case for single dish telescopes. Besides, \cite{Moore2015}
suggest that the foreground polarization fraction at $\sim$ 150 MHz is one order of magnitude
lower than that at 1.4 GHz.  Polarization calibration with single dishes is
potentially a challenge at 800 MHz. 

Performing polarization calibration with a signal dish radio telescope
with dual receptors, such as the GBT, usually entails determining the
Jones matrix, which describes the instrumental response of the two polarization
receptors to sky signals at each frequency channel. Polarized astrophysical
sources are often used to solve for the Jones matrices (e.g.:
\cite{Heiles2001} and \cite{Straten2004}). PSRCHIVE
(\cite{Hotan2004}), a software that is widely used for single dish
radio polarization calibration in the pulsar community, uses pulsar
observations at multiple
parallactic angles to solve for the Jones matrix at each frequency
channel. In this work, we use the Jones matrix model described by
\cite{Straten2004}, which is adopted in PSRCHIVE, to parametrize Jones
matrices.  We obtain both pulsar and quasar observations from the GBT.  
We first solve for the first-order approximation to the Jones matrix model parameters using pulsar data and the PSRCHIVE software, then fine tune the parameters using quasar observations.
We begin with a brief introduction to GBT 800 MHz receiver and the back-end systems we use in this work in Section \ref{sec:instrument}.
The polarization calibration on boresight is discussed in detail in Section \ref{sec:on_cal}.

Besides polarization leakage on boresight, another key element is to
characterize the polarized primary beam pattern of a telescope. 
Pulsar observations have the adventage that it is easy to separate on-axis pulsar signal and stationary off-axis
leakage by subtracting off-pulse from on-pulse data.
On the other hand, since both on-axis signal and off-axis leakage are
stationary in our intensity mapping observations, off-axis leakage
is therefore an issue that needs to be addressed.
In Section \ref{sec:beam}, we use spider scans of quasars and pulsars
to investigate the polarization beam pattern of the GBT. These
patterns are important for the interpretation of the HI intensity mapping power spectrum.
The procedure we are using to calibrate real data is summarized in Section \ref{sec:summary}.
We will discuss the results and limits of our investigation in Section \ref{sec:conclusion}.

\section{GBT 800 MHz receiver and GUPPI backend}\label{sec:instrument}
Here we briefly describe the GBT 800 MHz receiver and backend system.  The GBT has an off-axis optical design;  the 800 MHz receiver is a prime-focus instrument that operates at 680-920 MHz.  The feed is a corrugated feed horn with an Orthomode transducer (OMT) polarization splitter.  There are known resonances associated with the OMT at 796.6 MHz and 817.4 MHz, which we omit from the analysis.  A noise diode signal is injected after the OMT at 2 K level and switches at 15.26 Hz.  We use the Green Bank Ultimate Pulsar Processing Instrument (GUPPI) pulsar back-end systems \citep{GUPPI}, with a bandwidth of 200 MHz (700-900 MHz) over 4096 frequency channels, and integrate over 1 ms intervals. GUPPI has the pulsar-folding capability which we use in our analysis. See \cite{Masui2013} for more details.

\section{Polarization Calibration}\label{sec:on_cal}
In this section we present the method for polarization calibration at the beam
center, or boresight, without considering the angular response of the primary beam.  We adopt the
model described by \cite{Straten2004} to parametrize the Jones matrix
and use quasar and pulsar observations to find solutions for the Jones
matrix parameters.  We then apply the parameters on quasar data to
further correct for polarization leakage. 

\subsection{Mueller/Jones matrix model}\label{sec:model}
In the following sections we follow \cite{Britton2000} and
\cite{Straten2004} to model the Jones matrix.
The mapping between the real sky electric field signal $\vect{S}_\mathrm{E}=(E_{x},E_{y})$ 
and the observed signal $\vect{S}_\mathrm{E}^{'}=(E^{'}_{x},E^{'}_{y})$ through an
instrument is
\begin{equation}
\vect{S}_\mathrm{E}^{'}=\matr{JR_{\mathrm{E}}}(\phi)\vect{S}_\mathrm{E},
\label{eq:s'}
\end{equation}
where $\matr{J}$, which is typically called Jones matrix, is the instrumental response, and 
\begin{equation}
\matr{R_{\mathrm{E}}}(\phi)=\left(\begin{array}{cc}
 \cos\phi &\sin\phi\\
-\sin\phi &\cos\phi
\end{array}\right)
\end{equation}
is the relative rotation between the sky and the receptor by a
parallactic angle $\phi$. If we transform Eqn. (\ref{eq:s'}) into the basis of Stokes parameters, we get
\begin{equation}
\vect{S}^{'}_\mathrm{sp}=\matr{MR}(\phi)\vect{S}_\mathrm{sp},
\label{eq:s'sp}
\end{equation}
where $\vect{S}_\mathrm{sp}$ is a 4-vector representing Stokes ($I$, $Q$, $U$, $V$), $\matr{M}$ the $4\times4$ Mueller matrix, and
\begin{equation}
\matr{R}=\left(\begin{array}{cccc}
1&0&0&0\\       
0&\cos2\phi &\sin2\phi&0\\
0&-\sin2\phi &\cos2\phi&0\\
0&0&0&1
\end{array}\right).
\label{eq:rotation}
\end{equation}

Note the relation between the 2-vector $\vect{S}_\mathrm{E}$ and 4-vector $\vect{S}_\mathrm{sp}$ is
\begin{equation}
\vect{S}_\mathrm{sp}=\matr{A}(\vect{S}_\mathrm{E}\otimes \vect{S}_\mathrm{E}^{\ast}),
\end{equation}
where $\otimes$ is Kronecker product and 
\begin{equation}
\matr{A}=\left(\begin{array}{cccc}
1&0&0&1\\       
1&0&0&-1\\
0&1&1&0\\
0&i&-i&0
\end{array}\right).
\end{equation}

The relation between the Jones matrix $\matr{J}$ and Mueller matrix $\matr{M}$
is then 
\begin{equation}
\matr{M}=\matr{A}(\matr{J}\otimes \matr{J}^{\ast})\matr{A}^{-1}.
\label{eq:Jones_to_Mueller}
\end{equation}

$\matr{J}$ can be parametrized as
\begin{equation}
\matr{J}=G\matr{\Gamma}(\gamma)\matr{R_\mathrm{\Phi}}(\varphi)\matr{C},
\label{eq:jones}
\end{equation}
where $G$, the absolute gain, is a scalar. $\matr{\Gamma}(\gamma)$
is a matrix corresponding to the differential gain $\gamma$,
\begin{equation}
\matr{\Gamma}(\gamma)=\left( \begin{array}{cc}
e^{\gamma} & 0 \\
0          & e^{-\gamma}
\end{array}\right).
\end{equation}
$\matr{R_\mathrm{\Phi}}(\varphi)$ is a matrix corresponding to the differential phase $\varphi$,
\begin{equation}
\matr{R_\mathrm{\Phi}}(\varphi)=\left( \begin{array}{cc}
e^{i\varphi} & 0 \\
0          & e^{-i\varphi}
\end{array}\right).
\end{equation}
$\matr{C}$ is a matrix that represents a receiver with non-orthogonal receptors,
\begin{equation}
\matr{C}=\matr{\delta_{0}L}(\theta_{0},\epsilon_{0})+\matr{\delta_{1}L}(\theta_{1},\epsilon_{1}),
\end{equation}
where $\matr{L}(\theta,\epsilon)$ is written as
\begin{equation}
\matr{L}(\theta,\epsilon)=\left( \begin{array}{cc}
\cos \epsilon  & i\sin \epsilon \\
i\sin \epsilon & \cos \epsilon
\end{array}\right)
\left( \begin{array}{cc}
\cos \theta    &  \sin \theta \\
-\sin \theta & \cos \theta
\end{array}\right).
\end{equation}
$\epsilon$ and $\theta$ are the ellipticity and orientation of the
receptor, respectively.
$\matr{\delta_{a}}$ is the selection matrix,
\begin{equation}
\matr{\delta_{a}}=\left( \begin{array}{cc}
\delta_{0a} & 0 \\
0          & \delta_{1a}
\end{array}\right),
\end{equation}
where $\delta_{ab}$ is the Kronecker delta.

$\matr{C}$ can be decomposed as
\begin{equation}
\matr{C}(\epsilon_{0},\theta_{0},\epsilon_{1},\theta_{1})=\matr{C}(\epsilon_{0},\theta_{0}-\phi_{0},\epsilon_{1},\theta_{1}-\phi_{0})\matr{R_{\mathrm{E}}}(\phi_{0}),
\label{eq:C_decompose1}
\end{equation} 
where
\begin{equation}
\matr{R_{\mathrm{E}}}(\phi_{0})=\left(\begin{array}{cc}
 \cos\phi_{0}&\sin\phi_{0}\\
-\sin\phi_{0}&\cos\phi_{0}
\end{array}\right)
\label{eq:C_decompose2}
\end{equation}
is equivalent to the parallactic angle rotation. Therefore, one can define $\theta_{+}\equiv\theta_{0}+\theta_{1}$ and $\theta_{-}\equiv\theta_{0}-\theta_{1}$ to separate the overall rotation from the relative rotation between receptors, and merge $\theta_{+}$ into parallactic angle rotation.

Then one can transform $\matr{J}$ to Mueller matrix $\matr{M}$ with Eqn. (\ref{eq:Jones_to_Mueller}).

To illustrate some of the structure of the Jones and Mueller matrices, we set ($\theta_{0}, \theta_{1}, \epsilon_{0}, \epsilon_{1})=0$, Eqn. (\ref{eq:jones}) can be rewritten as:
\begin{equation}
\matr{J}=G\left( \begin{array}{cc}
e^{\gamma +i\varphi} & 0 \\
0          & e^{-\gamma -i\varphi}
\end{array}\right),
\label{eq:ncal_jones}
\end{equation}
and the Mueller matrix becomes
\begin{equation}
\matr{M_{0}}=G^{2}\left( \begin{array}{cccc}
\cosh(2\gamma) & \sinh(2\gamma) &   0    &    0    \\
\sinh(2\gamma) & \cosh(2\gamma) &   0    &    0    \\
           0         &        0     &    \cos{2\varphi}     &   \sin{2\varphi}       \\
           0         &        0     &   -\sin{2\varphi}     &   \cos{2\varphi}       \\
\end{array}\right).
\label{eq:ncal_Mueller}
\end{equation}

In summary, the Jones matrix is a $2\times 2$ complex matrix, in principle it
has eight degrees of freedom.
However, from Eqn. (\ref{eq:Jones_to_Mueller}), one can see that the
absolute phase of $\matr{J}$ does not affect the Mueller matrix $\matr{M}$. Therefore, the complete
Jones/Mueller matrix model can be described by seven parameters, which
are the absolute gain $G$, the differential gain $\gamma$, the differential phase $\varphi$, $\theta_{0}, \theta_{1}$ the orientation of
the two receptors, and the ellipticity of the receptors $\epsilon_{0},\epsilon_{1}$.

\subsection{Polarization calibration with pulsar\label{sec:pulsar}}
\subsubsection{Solving Mueller matrix with Pulsar data}
One can observe a polarized source, such as a pulsar, to measure
$\vect{S}^{'}_\mathrm{E}$ at parallactic angle $\phi$. With data taken at several
parallactic angles, one will have a set of simultaneous equations,
$\vect{S}^{'}_{\mathrm{E}i}=\matr{JR_{\mathrm{E}}}(\phi_{i})\vect{S}_\mathrm{E}$, or
\begin{equation}
\vect{S}^{'}_{\mathrm{sp}i}=\matr{MR}(\phi_{i})\vect{S}_\mathrm{sp}
\label{eq:multi-angle}
\end{equation} in the Stokes parameter basis. $i$ indexes the observation at parallactic angle $\phi_i$

Before solving for both the incoming signals of the
pulsar and the Jones matrix parameters using these equations, we have to deal with two
degeneracies: One of the degeneracies is between the Stokes parameters
$I$ and $V$ of the incoming signal, and the other is between Stokes $Q$ and
$U$. 

The degeneracies arise from the following: In
Eqn. (\ref{eq:multi-angle}), one can equally substitute $\matr{M}$ and $\vect{S}_\mathrm{sp}$ with 
$\matr{MD}^{-1}$ and $\matr{D}\vect{S}_\mathrm{sp}$, respectively, as long as the matrix $\matr{D}$ commutes with $\matr{R}$.  In other words, if $\vect{S}_\mathrm{sp}$ is
one solution of Eqn. (\ref{eq:multi-angle}), $\matr{D}\vect{S}_\mathrm{sp}$ will be
another solution as well.  In fact, any matrix $\matr{D}$ of the following form,
\begin{equation}
\matr{D}=a\left(\begin{array}{cccc}
1&0&0&V_{1}\\       
0&1&0&0\\
0&0&1&0\\
V_{2}&0&0&1
\end{array}\right)
+
b\left(\begin{array}{cccc}
1&0&0&0\\       
0&\cos\theta &\sin\theta &0\\
0&-\sin\theta &\cos\theta &0\\
0&0&0&1
\end{array}\right),
\label{eq:degeneracy}
\end{equation}
will commute with $\matr{R}$ (see Appendix B of \cite{Straten2004}). The first term of Eqn. (\ref{eq:degeneracy}) corresponds to degeneracy between $I$ and $V$, and the second term degeneracy between $Q$ and $U$. 

One way to break the degeneracy between Stokes $I$ and $V$ is to observe a
standard calibrator with a known ratio of Stokes $I$ to $V$.  There is a
built-in noise diode at the GBT that can serve this function. Ideally,
the noise diode will produce pure linearly polarized signals at a
position angle $45^{\circ}$ to the two orthogonal receptors. We can
therefore break the degeneracy between Stokes $I$ and $V$ by assuming that
the noise diode produces no circular polarization
signals. Alternatively, we can assume the system circular
  polarization $V$ to be negligible while observing astrophysical calibrators.

As is the standard procedure, by taking the differences of on-
and off-source observations of a standard calibrator, such as 3C295,
one can estimate its circular polarization. One can therefore break the degeneracy by assuming the Stokes 
$V$ of standard calibrators, which are reported to have a negligible level of $V$, to be zero.

The degeneracy between $Q$ and $U$, which makes it impossible to calibrate absolute polarization angles on the sky without an external reference, is related to the definition of angles. 
Here we set the orientation of the first receptor by fixing $\theta_{0}=0$. Therefore, $\theta_{-} = -\theta_{1}$ in the following sections.

PSRCHIVE solves $\matr{J}$ by using a least-squares minimization method
for each frequency channel (see \cite{Straten2004} for more
detail). It automatically selects several pulsar pulse phase bins,
from pulsar data taken at different parallactic angles, to solve for
the model parameters. The unknown variables are the Stokes parameters of the
incoming pulsar signals and the Jones matrix parameters of the
instrument.  Because the total flux density $I$ of pulsars typically
varies on a time-scale of minutes, PSRCHIVE normalizes the Stokes
parameters by the invariant interval $S=I^2-Q^2-U^2-V^2$ instead (details in \cite{Straten2004}).
We can therefore compare pulsar Stokes parameters between different scans.

However, the PSRCHIVE does not directly assume the circular polarization of the standard candle to be zero.
Instead, PSRCHIVE provides two options: either assume the $V$ of system temperature while directing the telescope to a standard candle
, or the $V$ of noise diode, to be zero.
Following \cite{Straten2004}, who suggests that the circular polarization of a
noise diode is significantly different from zero, we choose the first option.

With pulsar observations at five different parallactic angles, we 
have 20 measurements (4 polarizations $\times$ 5 parallactic angles)
at each frequency channel and each selected pulse phase bin. In
addition to the flux calibrator data from quasar observations (four
polarizations) and the noise diode data (four polarizations), we have
328 measurements for 16 selected pulse phase bins at each frequency
channel. The number of variables at each frequency is 76, including a
total of 64 Stokes parameters of the pulsar for the 16 selected pulse
phase bins, six Stokes parameters of the flux calibrator and noise
diode (the $I$ of flux calibrator is assumed to be known, and the $V$
of flux calibrator is set to be zero), and six instrumental
Jones/Mueller matrix model parameters (i.e., the seven parameters
describe in Section \ref{sec:model} except for $\theta_{0}$, which is set
to be $0$ to break the degeneracy between $Q$ and $U$). Once PSRCHIVE
obtains the best fit instrumental parameters, it generates both the
Jones matrix and Mueller matrix for each frequency channel with the
formalism described in the above section.

\begin{figure}
\plotone{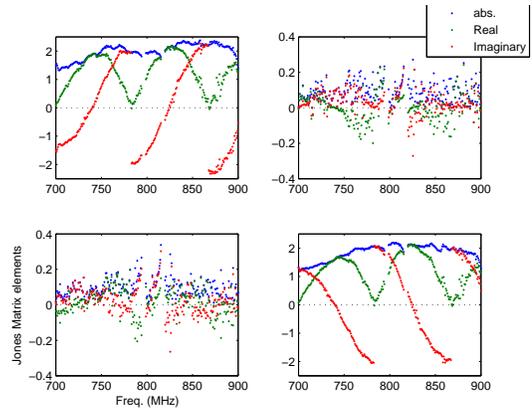}
\caption{ The Jones matrix elements of GBT as a function of frequency. The upper left is $\matr{J}(0,0)$, upper right $\matr{J}(0,1)$, lower left $\matr{J}(1,0)$, and lower right $\matr{J}(1,1)$. The Jones matrix is obtained from B1133+16 data. The
  unit is arbitrary.   The features nears frequencies 800 MHz and
  820 MHz are due to the resonance in the
  orthomode transducer (OMT) in this band at the GBT, and are excluded
  from all analysis in this paper.\label{Jones_a_B1133}}
\end{figure}

\begin{figure}
\plotone{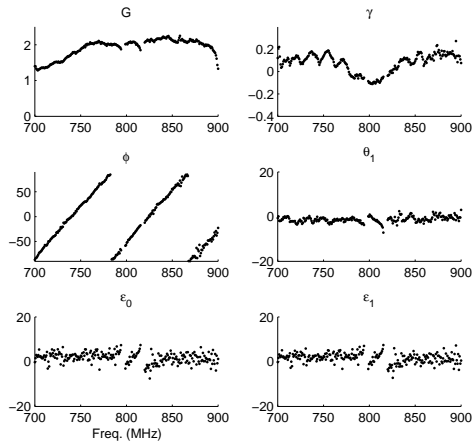}
\caption{The values of Jones matrix parameters of GBT as a function of frequency. The parameters are obtained from B1133+16 data. The unit of $G$ is abitrary, $\gamma$ is dimensionless. $\varphi$, $\theta_{1}$, $\epsilon_{0}$, and $\epsilon_{1}$ are in units of degree.\label{fig:Jones_p_B1133}}
\end{figure}

Figures \ref{Jones_a_B1133} and \ref{fig:Jones_p_B1133} show the Jones
matrix obtained from pulsar data, where we took five five-minute
tracking scans for pulsar
B1133+16 in Jan, 2014 at different parallactic angles with the
GBT. The standard candle used is quasar 3C295, which is known to be an unpolarized source at 800 MHz.
The model parameters shown in Figure \ref{fig:Jones_p_B1133} are consistent with the ones shown by \cite{Han2009} in the overall.

\subsubsection{Determining parameters with noise diode \label{sec:G_gamma_varphi}}
We find that at the GBT, the value of $\varphi$ changes dramatically day by day and
needs to be determined with each observation;  the values of $G$ and
$\gamma$ are also found to change slightly day by day and
significantly several times in a observing semester over six months.

Since at the GBT, the noise diode signal is injected into the
receiver only after the separation of orthogonal polarizations, parameters
$\epsilon_{0,1}$ and $\theta_{0,1}$, which describe the cross-talk
between orthogonal polarizations, are expected to be negligible for noise diode signal. The
Mueller matrix that applies to the noise diode signal can thus be
written as Eqn. (\ref{eq:ncal_Mueller}).  

One can therefore solve for $G$, $\gamma$, and $\varphi$ by inserting a known noise
diode standard profile $\vect{S}_\mathrm{ncal}$ and the observed noise diode data
$\vect{S}^{'}_\mathrm{ncal}$ into
\begin{equation}
\matr{M_{0}}\vect{S}_\mathrm{ncal}=\vect{S^{'}}_\mathrm{ncal}.
\label{eq:M0_ncal}
\end{equation}
 Assuming the noise diode frequency profile is
stable over a few months, we can use the noise diode data calibrated
by PSRCHIVE as the standard profile to solve for $G$, $\gamma$, and
$\varphi$ for each observation session. Combining these parameters
with $\theta_1$, $\epsilon_0$, and $\epsilon_1$ obtained from multiple
parallactic-angle tracking scans of B1133+16, we can achieve $\sim 1.3\%$
accuracy on all the Stokes parameters, with respect to total intensity, of the
unpolarized quasar 3C147. The accuracy of $1.3\%$ is the RMS
fluctuation of dozens of calibrated tracking scans on 3C147 over an entire semester.

\subsection{Parameter modification with Quasar data}

\subsubsection{Previous Assumptions}

The calibration method described above is based on the following
assumption: The linear polarization fraction, angle, and the circular
polarization fraction of the target pulsar are assumed to be stable
over the observation period, which is about six hours. To test the
first assumption, we examine the calibrated pulsar properties as a function
of time. Figure \ref{fig:RMS_p_f_B1133} shows the RMS of linear
polarization fraction over five calibrated pulsar tracking scans taken in the
same night. We find that the RMS fluctuation of the linear polarization fraction measured over
multiple parallactic-angle pulsar tracking scans is higher than the thermal noise level. It appears the polarization
fraction of pulsars is not stable enough for our purpose. The RMS
fluctuation of the linear polarization angle ($\approx
2^{\circ}$-$4^{\circ}$) and circular polarization fraction ($\approx 1\%$-$3\%$) are also
larger than acceptable.  

\begin{figure}
\plotone{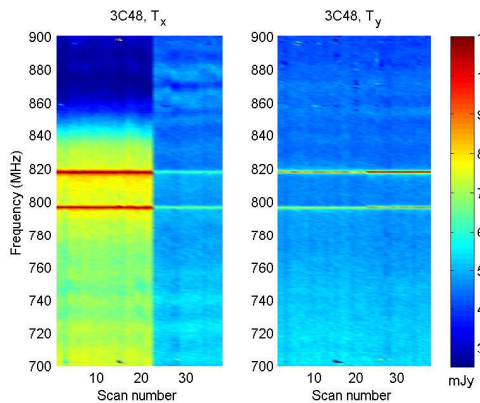}
\caption{An example of instability of GBT noise diode. In this figure
  we show the $T_x$ and $T_y$ of noise diode derived by comparing
  noise diode data with 3C48 data, which were taken at the same
  time. The left panel and right panel are $T_x$ and $T_y$ of noise
  diode, respectively. The intensity unit here is mJy. The dataset
  shown here consists 38 tracking scans spreading from May to August,
  2011. The two horizontal stripes at frequencies near 800 MHz and
  820 MHz correspond to bad channels due to the resonance in the
  orthomode transducer (OMT) in this band at the GBT, and are excluded from all analysis. \label{fig:Tcal}}
\end{figure}

Because it takes a good amount of observing time (including
overhead) to conduct multiple
parallactic angle pulsar tracking scans, during our HI intensity mapping
observing campaign spread over a few semesters, we only take multiple
parallactic-angle pulsar tracking scans once every several months. The
stability of the noise diode profile becomes crucial. The stability is
tested with observations of on- and off-source quasar tracking scans while
blinking the thermal noise diode at 15.26 Hz in every observing
session.  Comparing data with and without a bright quasar in the
center of the beam, and data with and without the noise diode signal, one
can derive $T_{x}$ and $T_{y}$, the injected noise diode temperature
in the two polarization directions, given the known
spectrum of the quasar.  We find that the noise diode is not stable
over a time span of six months either. Figure
\ref{fig:Tcal} shows noise diode temperature measured with scans on
3C48 in 2011.  There appears to be a discontinuity in noise diode
temperature at scan 22, below which the receptor X has a higher
temperature than receptor Y but with a 2dB attenuation above 840
MHz. This discontinuity was observed between June 27, 2011 and July
17, 2011, which coincided with the maintenance activity at the
GBT when the 800 MHz receiver was taken down from the prime focus.  It
is quite possible that the noise diode was reset when the
receiver reinstalled. Therefore, we need to modify the method of estimating
the Jones/Mueller matrix for the GBT.

\subsubsection{Noise Diode Behavior\label{sec:ncal_profile}}

In order to reduce the impact of noise diode instability, we examine $T_{x}$ and $T_{y}$ as a function of time and frequency, derived from quasar tracking scans in each session,
then group all sessions into subsets according to the shapes of the
derived noise diode frequency profiles. In practice, the number of
identified noise diode profile shapes in each semester varies between 1 to 9 for the five semesters
from 2011 to 2015. We average the $T_{x}(\nu)$ and $T_{y}(\nu)$ profiles within each
subset to obtain $\tilde{T}_{x,j}(\nu)$ and $\tilde{T}_{y,j}(\nu)$, where $j$ indexes subsets. Then we
normalize the profiles to $\left
  \langle\sqrt{\tilde{T}_{x,j}\tilde{T}_{y,j}}\right\rangle_{\nu}=1$, averaging over all frequency channels. We further average $\left\langle\sqrt{T_{x}T_{y}}\right\rangle_{\nu}$ over all tracking scans within each session to get a
normalization factor $N_{k}$ for each session, where $k$ indexes sessions. Finally we obtain estimated noise
diode profile for each session as $N_{k}\tilde{T}_{x,j}$ and
$N_{k}\tilde{T}_{y,j}$, then apply them to correct $\vect{S}_\mathrm{ncal}$ in Eqn. (\ref{eq:M0_ncal}). In order to reduce the impact of errors associated with the estimation
of noise diode profiles, we only allow $\langle T_{x}T_{y}\rangle_{\nu}$ to vary between
sessions, and fix the profile shape within each subset. 

\subsubsection{First-order Correction\label{sec:1st_order}}

Assuming $\theta_{1},\epsilon_{0},\epsilon_{1}\ll 1$, to first-order
approximation we can calculate the resulting calibration errors induced 
by errors in values of Jones matrix parameters using Eqn. (\ref{eq:S''}) and
(\ref{eq:RdeltaMR}).  The formalism is described in detail in Appendix A.  Figure \ref{fig:Jones_p_B1133} suggests that the assumption
of small $\theta_{1}$, $\epsilon_{0}$, and $\epsilon_{1}$ values is
sound for the GBT. We can also define $\epsilon_{+}\equiv\epsilon_{0}+\epsilon_{1}$ and $\epsilon_{-}\equiv\epsilon_{0}-\epsilon_{1}$,
for they are directly associate with $Q$-$V$ and $I$-$V$ leakages, respectively (see Appendix A and \cite{Britton2000}).

\begin{figure}
\plotone{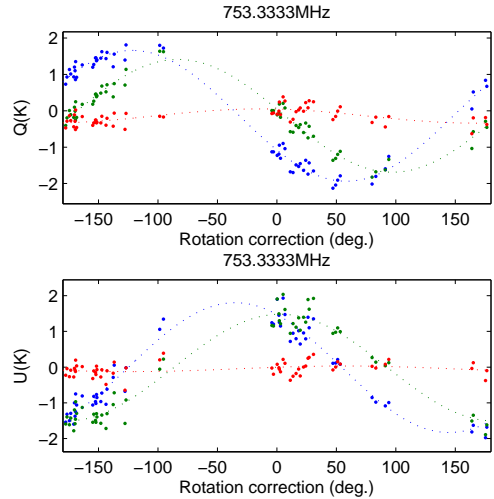}
\caption{Comparison of calibrated $Q$ (upper) and $U$ (lower) of an unpolarized source 3C295 with/without parameter modifications. The blue dots show the data calibrated by the method described in Section \ref{sec:pulsar}, the green dots are calibrated with modified $\gamma$, and the red ones are calibrated with modified $\gamma$ and $\theta_{-}$. Each dot corresponds to one 1-minute tracking on 3C295. The parameter modification is described in Section \ref{sec:1st_order}. The rotation correction here is defined as $2\phi$. The total intensity of 3C295 in frequency range 700 - 900 MHz spread from 58 K to 72 K.\label{fig:mod_gamma_theta_1}}
\end{figure}

For an unpolarized source like quasar 3C295, one can apply $\vect{S\matr{_{sp}}}=[I, 0, 0, 0]$ to Eqn. (\ref{eq:S''}) and (\ref{eq:RdeltaMR}), and get
\begin{equation}
\delta Q/I=-(2\delta\gamma\cos2\phi-\delta\theta_{-}\sin2\phi)
\label{eq:deltaQ}
\end{equation} 
and
\begin{equation}
\delta U/I=-\delta\theta_{-}\cos2\phi-2\delta\gamma\sin2\phi. 
\label{eq:deltaU}
\end{equation} 
These equations predict a sinusoidal dependence of the calibrated $Q$
and $U$ on parallactic angles.  The predicted sinusoidal pattern is
shown as blue dots in Figure \ref{fig:mod_gamma_theta_1}, which represent the $Q$ and $U$ values of 3C295 data calibrated with a Mueller matrix calculated from pulsar observations,

For an unpolarized source, 
the calibrated values of $Q$ and $U$, which are otherwise expected to
be zero with perfect polarization calibration, are denoted as $\delta Q$ and $\delta U$.
Substituting these values into Eqns. (\ref{eq:deltaQ}) and
(\ref{eq:deltaU}), we obtain the best-fit values of $\delta\gamma$ and $\delta\theta_{-}$.

As a consistency check, we calculate the values of $\delta \gamma$ and
$\delta \theta_{-}$ with data on 3C147, adjust the $\gamma$ and
$\theta_{-}$ parameters accordingly and apply them to the polarization
calibration of 3C295 data. The 3C295 data calibrated with the modified
parameters are shown as green and red dots in Figure
\ref{fig:mod_gamma_theta_1}. Modification of $\gamma$ and $\theta_{-}$
dramatically reduce the fluctuation of calibrated $Q$ and $U$ of
3C295. This implies the modification of $\gamma$ and $\theta_{-}$
parameters is a sound approach. 

\begin{figure}
\plotone{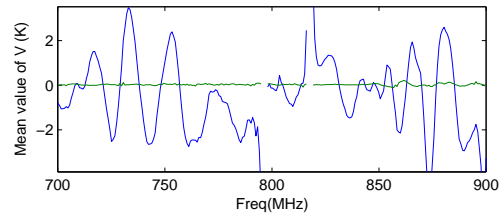}
\caption{Calibrated $V$ of 3C295. The blue curve shows the calibrated
  $V$ from pulsar tracking scans, and the green curve includes the $\epsilon_{-}$
  correction with data from 3C147, which helps substantially in
  achieving $V \sim 0$. The total intensity of 3C295 in frequency range 700 - 900 MHz spread from 58 K to 72 K.\label{fig:mod_eps_minus}}
\end{figure}

On the other hand, the matrix element $(4,1)$ in the right hand side of Eqn. (\ref{eq:RdeltaMR}) describes the leakage from $I$ into $V$ due to $\delta\epsilon_{-}$. The circular polarization level of 3C295 is smaller than $0.6\%$ at our frequency band, thus we assume that the oscillating pattern of $V$ of calibrated 3C295 data, which is shown as the blue curve in Figure \ref{fig:mod_eps_minus}, is caused by $\delta\epsilon_{-}$ in the parameter set we applied. Just as for $\delta\theta_{-}$ and $\delta\gamma$, we also use a modified $\epsilon_{-}$ derived from 3C147 data to calibrate 3C295 data. The corrected $V$ of 3C295 is much closer to zero as shown by the green curve in Figure \ref{fig:mod_eps_minus}.

\begin{figure*}
\plotone{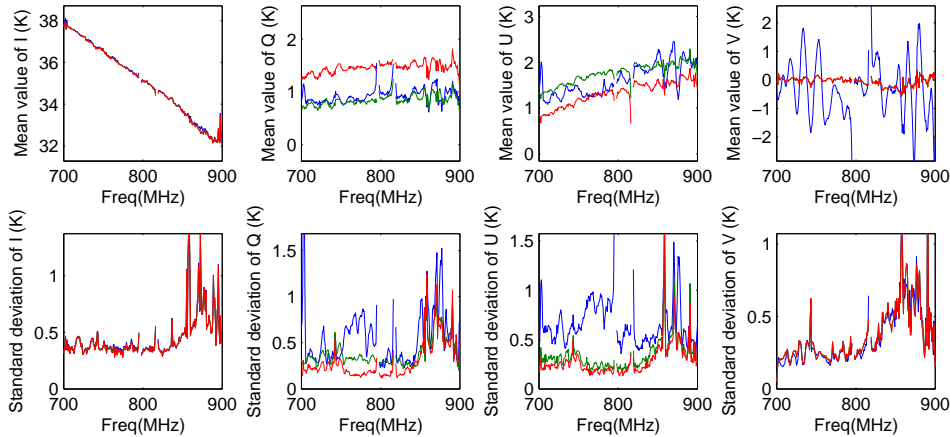}
\caption{Calibrated 3C286 data. The upper panels show calibrated Stokes parameters averaged over 57 tracking scans on 3C286, while the lower ones show the standard deviations between those tracking scans. The blue lines show the result of calibration with parameters solved from pulsar data, the green lines correspond to calibration with modified $\theta_{-}$ and $\epsilon_{-}$ (based on 3C48 data), and the calibration shown as red lines using exactly same parameters except additional ionospheric Rotation Measure correction.  \label{fig:cal_compare_A}}
\end{figure*}

Once we obtain the modified $\gamma$, $\theta_{-}$, and $\epsilon_{-}$
from quasar data, one can apply the solutions to other quasars
including 3C48, 3C295, 3C147, and 3C286. The RMS of calibrated stokes
parameters over dozens of tracking scans through an entire observing semester
are about $0.6\%$-$0.8\%$ of the quasar intensity in most of the
frequency channels. 3C286, a slightly polarized quasar, is taken as an
example to illustrate the outcome of the parameter correction, indicated by the blue and green lines in Figure \ref{fig:cal_compare_A}.

The modified $\theta_{-}$ can be used to test the assumption
that the cross talk between $T_x$ and $T_y$ of the noise diode is negligible.
As shown in Eqn. (\ref{eq:deltaM}), $\theta_{-}$ describes the leakage between $I$ and $U$.
The $\theta_{-}$ experienced by the sky signal can be approximated as
 a linear combination of 
$\theta_{-}$ before and after noise diode signal injection. We denote the latter one as $\theta_{-,\mathrm{ncal}}$ because it is the $\theta{-}$ which noise diode signal experienced. 
 As $U_{\mathrm{ncal}}$ is strong ($\approx 80\%$ of
$I_{\mathrm{ncal}}$), fluctuations in $\theta_{-,\mathrm{ncal}}$
will contribute to the fluctuations in $I_{\mathrm{ncal}}$.
Therefore, we postulate that the measured $I_{\mathrm{ncal}}$ is to correlate with the
$\theta_{-}$ value estimated from quasar data, if the fluctuation of $\theta_{-,\mathrm{ncal}}$ is not negligible.

We therefore correlate fluctuations over time and frequency channels 
of $I_{\mathrm{ncal}}$ and $\theta_{-}$ obtained from quasar 3C48
taken in 2014 and 2015.
The most prominent patterns extracted using
singular-value decomposition in frequency-time space are removed before correlation. These removed
patterns of $I_{\mathrm{ncal}}$ and $\theta_{-}$ do not correlate with
each other. On the other hand, from previous noise diode data, we found that $T_{x,\mathrm{ncal}}$ and $T_{y,\mathrm{ncal}}$
tend to preserve spectral shape over several nights with normalization changes each night.
This kind of pattern matches what we get in the first SVD mode.
These two facts imply that these removed modes can be considered to represent fluctuations of
$I_{\mathrm{ncal}}$ and $\theta_{-}$ before the light path of sky
signal merge with the noise diode signal. 
The residual $I_{\mathrm{ncal}}$ and $\theta_{-}$ have a correlation
coefficient of $r=0.24\pm0.03$, significantly different
from $0$.  As a sanity check, we remove the correlated part from
$I_{\mathrm{ncal}}$, and find the RMS of measured $I_{\mathrm{ncal}}$
only reduced by $\approx 0.04\%$ of total $I_{\mathrm{ncal}}$.
Although there is statistically significant correlation between residual $I_{\mathrm{ncal}}$ and $\theta_{-}$, the impact of this correlation is much smaller than the uncertainty level, $0.6\%$-$0.8\%$, of our calibration.  
We perform a similar check on $\epsilon_{-}$ and noise diode
$V_{\mathrm{ncal}}$, for $\delta\epsilon_{-}$ would results in leakage from $I_{\mathrm{ncal}}$ into $V_{\mathrm{ncal}}$
if $\delta\epsilon_{-}$ happens after noise diode signal injecting into the receiver.
We do not find significant correlation between $\epsilon_{-}$ and the $V_{\mathrm{ncal}}$.
We therefore conclude that the cross talk between $T_{x,\mathrm{ncal}}$ and $T_{y,\mathrm{ncal}}$ is negligible.

\subsection{Ionospheric Rotation Measure Correction\label{sec:RM}}

Ionospheric Faraday rotation is another effect that needs to be
accounted for in the data. Faraday rotation is the rotation of linear
polarization direction when photons propagate through magnetized free
electrons. The angle of rotation is proportional to the wavelength $\lambda$
of light as $\lambda^{2}$, and can be written as $\theta_\mathrm{FR}=\phi_\mathrm{RM}\lambda^{2}$, where $\phi_\mathrm{RM}$ is Rotation Measure (RM). RM can be calculated as
\begin{equation}
\phi_\mathrm{RM}\approx(2.62\times 10^{-13} \mathrm{T}^{-1})\int n_\mathrm{e}(s)B_{\parallel}(s)ds,
\label{eq:RM}
\end{equation}
where $n_\mathrm{e}$ is free electron density, $B_{\parallel}$ is the strength of magnetic
field component parallel to the line of sight, T stands for Tesla, the SI unit magnetic
field strength, and $ds$ is integrated along the
line of sight. 

Photons from astronomical sources propagate through the terrestrial ionosphere, composed of the free electrons
magnetized by the geomagnetic field, the linear polarization angle of
light is thus rotated by the Faraday effect. The RM depends on the spatial distribution of $n_{e}$, which is a function of time, and the trajectory of light propagation, which is determined by azimuth and elevation of the pointing of the telescope.  We adopt empirical orthonormal functions (EOFs)\citep{USTEC0} and incorporate ionospheric information released by US-TEC (\cite{USTEC0} and \cite{USTEC}) to estimate free electron
density at the time of observation. The geomagnetic field is calculated using the International
Geomagnetic Reference Field (IGRF) \citep{Finlay2010}.    After
applying the free electron number density and geomagnetic field to
Eqn. (\ref{eq:RM}), we obtain the RM values and calculate $\theta_\mathrm{FR}$ caused by ionospheric free electrons. Faraday rotation can be corrected by simply adding $\theta_\mathrm{FR}$ to the parallactic rotation angle $\phi$ in Eqn. (\ref{eq:rotation}). 

With $\phi_\mathrm{RM}=2m^{-2}$, which is a typical ionospheric $\phi_\mathrm{RM}$ value
at the GBT site during night time when we observe, $\theta_\mathrm{FR}$ is
$12.7^{\circ}$ at 900 MHz and $21.0^{\circ}$ at 700 MHz. The ionospheric
Faraday rotation affects the linear polarization direction rather
significantly in our observing frequency range.

\begin{figure}
\plotone{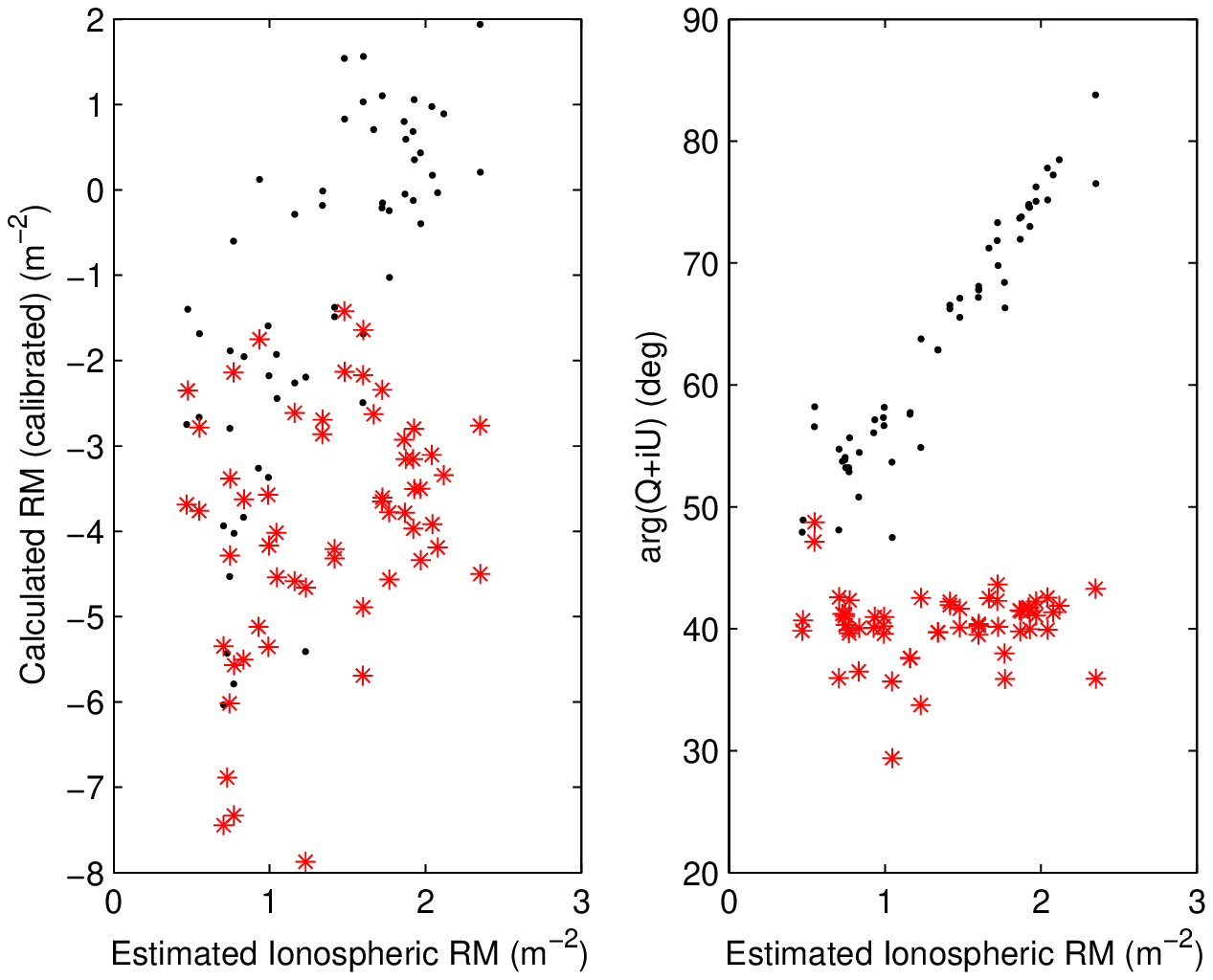}
\caption{Comparison of linear polarization angle of 3C286 with (red asterisks) and without (black dots) ionospheric RM correction.
The left panel shows the RM values fitted as the slope of linear
  polarization angles with respect to $\lambda^{2}$ using calibrated 3C286 data.     
  The right panel are the phase angles of calibrated $Q+iU$ averaged over frequency channels. The
  x-axes are ionospheric RM values estimated with US-TEC model. 
  The correlation coefficient $r$ between x-axis and y-axis values of black dots, which show the results before RM correction, is $0.73$ in the left panel and $0.95$ in the right one. The correlation coefficients become $0.29$ and $0.02$, respectively, after applying RM correction. 
  As shown here, the RM correction significantly
    reduces the variance and the averaged values of polarization
    angles of 3C286. }  \label{fig:RM_corrected}
\end{figure}

Figure \ref{fig:RM_corrected} shows the comparison of RM values
fitted as the slope of linear polarization angles with respect to
$\lambda^{2}$ using calibrated 3C286 data, the phase angles of calibrated 3C286 data averaged over frequency channels, and estimated
ionospheric RM.
The right panel of Figure \ref{fig:RM_corrected} shows that our RM correction effectively reduces the fluctuation
of polarization angle over time.

In principle, the remaining RM after ionospheric RM correction should
be the RM of the celestial object.  However, it is reported that the
RM of 3C286 is very close to zero \citep{Perley2013}, which
significantly differs from what we measure in Figure \ref{fig:RM_corrected}.
Note though that the FWHM of the GBT beam in our frequency band is about
$15^{'}-19^{'}$, the on-source 3C286 data will thus include contributions
from the surrounding Galactic Foreground.  Although we subtract
off-source data, which is taken 3 degrees away from 3C286, the
Galactic foreground at these two patches may differ and affect the
estimated RM value.

We have therefore taken 33 tracking scans of pulsar B1929+10.  We eliminate Galactic foreground
RM contribution by gating the pulsar and subtracting off-pulse from
on-pulse data, and expect the pulsar RM measurement to be free of
Galactic foreground contamination.  After ionospheric RM correction, the pulsar RM calculated
from the slope of $arg(Q+iU)$ changes from $-5.9 \pm 0.3 m^{-2}$ to
$-7.1 \pm 0.3 m^{-2}$, which is
consistent with the RM value from the literature, $-6.9 m^{-2}$ \citep{Johnston2005}.
We therefore validate our ionospheric RM correction procedure.

\begin{figure}
\plotone{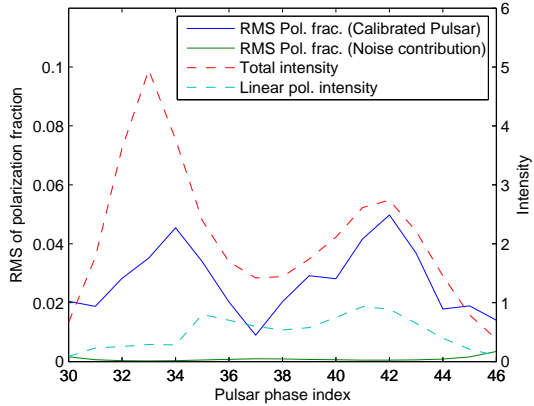}
\caption{The RMS of linear polarization fraction of pulsar B1133+16. The X axis is pulsar phase index. The blue curve is derived by comparing 5 tracking scans on B1133+16 taken at the same night. The green curve shows the contribution of noise. The dashed lines show the total and linear polarization intensity of the pulsar. The unit of intensities is arbitrarily. \label{fig:RMS_p_f_B1133}} 
\end{figure} 

\section{Polarized Beam Response}\label{sec:beam}

In the previous section we have focused on on-axis (boresight) polarization
calibration only.  However, off-axis polarization leakage can
potentially contaminate HI intensity maps and power spectrum
estimation. Considering the full polarized beam response, Eqn.
(\ref{eq:s'sp}) should be rewritten as a convolution,
\begin{equation}
\vect{S'}(\vec{x})=\int \matr{M}(\Delta\vec{x})\matr{R}(\phi)\vect{S}(\vec{x'})d^{2}x',
\label{eq:s'beam}
\end{equation}
where $\vec{x}$ and $\vec{x'}$ are the two-dimensional positions in
the sky, $\Delta\vec{x}=\vec{x'}-\vec{x}$. $\vec{x}$ and
    $\vec{x'}$ are defined in horizontal coordinate system. $\vec{x}$
    is the position of the boresight and $\vec{x'}$ the position of
    the source on the sky, so that $\Delta\vec{x}$ represents the
    source position relative to the boresight on the sky.
 In the case of a
point source, where $\vect{S}(\vec{x'})=\vect{S_{0}}\delta(\vec{x'}-\vec{x_{0}})$,
Eqn. (\ref{eq:s'beam}) reduces to Eqn. (\ref{eq:s'sp}) with
$\matr{M}=\matr{M}(\vec{x_0}-\vec{x})$. In this section we use quasars and pulsars
as probes to investigate the feature of $\matr{M}(\Delta\vec{x})$. We also
discuss the advantages and weaknesses of these two types of targets
for polarization calibration.

\subsection{Polarized Beam from Quasar spider scans}

\subsubsection{Polarized Beam Pattern}\label{sec:pol_beam_pattern}

We perform the so-called `spider scans' on unpolarized quasars to
investigate the structure of the first column of the Mueller matrix
$\matr{M}(\Delta\vec{x})$, i.e., the leakage contributions from intensity to
polarization.  To conduct a spider scan, we slew the telescope along four
$1^{\circ}$ paths across the
central source:  one of the four paths is iso-azimuthal, one
 iso-elevation, and the other two are $45^{\circ}$ from the first
two (shown as black lines in the upper panels of Figure
\ref{fig:spider_scan_map}). The lengths of the paths are $4.0$ and $3.2$ times of the beam FWHM at 900 MHz and 700 MHz, respectively.
We select a few unpolarized quasars,
including 3C48, 3C295, and 3C147, as targets. The upper panels of
Figure \ref{fig:spider_scan_map} show the beam maps from direct
interpolation of 3C295 spider scans at a single frequency channel. The
data shown in Figure \ref{fig:spider_scan_map} have been calibrated with the boresight calibration method described in Section \ref{sec:on_cal}.

In order to quantify the resulted polarization beams and minimize the impact
of observational noise, we model the spider scan maps with
the two-dimensional Gauss-Hermite functions, which are 
perturbations around a two-dimensional Gaussian profile.  The 2D
Gauss-Hermite function is written as
\begin{equation}
\mu_{ij}\left(\frac{x}{x_{0}},\frac{y}{y_{0}}\right)=e^{\frac{-(x^{2}+y^{2})}{2\sigma^{2}}}H_{i}\left(\frac{x}{x_{0}}\right)H_{j}\left(\frac{y}{y_{0}}\right),
\label{eq:gaussian_hermite_function}
\end{equation}
where the $\sigma$ is the beam size, $H$ the Hermite polynomials,
$x_{0}$ and $y_{0}$ are characteristic scales in the $x$ and $y$ directions.

The Gauss-Hermite fit is then
\begin{equation}
S_{(I,Q,U,V)}(x,y)=\sum^{n}_{i=0}\sum^{n}_{j=0}a_{(I,Q,U,V),ij}\mu_{ij}\left(\frac{x}{x_{0}},\frac{y}{y_{0}}\right),
\label{eq:GH_fitting}
\end{equation}
where $a_{(I,Q,U,V),ij}$ are fitting coefficients.

We choose $n=2$ to avoid over-fitting. $\sigma$ is set to be the
best-fit Gaussian beam size $\sigma_g$ of $I$, and $x_0=y_0=\sigma_g$. 
The lower panels of Figure \ref{fig:spider_scan_map} show the best fit $S_{I,Q,U,V}$. 

The most prominent feature of the derived polarized beam patterns is
the dipole shape of the Stokes $V$ beam. The dipole peak and trough are
$\approx 12\%$ of the intensity at beam center.  The dipole pattern in
$V$ is expected due to the off-axis design of the GBT \citep{Srikanth2012}.

\subsubsection{Q-V and U-V coupling}\label{sec:QUV}

We further discover that the derived $Q$ and $U$ beam patterns show
similar dipole shapes as the $V$ beam, albeit at a lower level. Eqn. (\ref{eq:deltaM})
indicates that the errors in $\epsilon_{+}$ and $\varphi$ may induce
leakage from $V$ to $Q$ and $U$, respectively. Eqn. (\ref{eq:deltaM}) also
implies that $\epsilon_{+}$ and $\varphi$ are not easy to be
constrained solely from unpolarized sources because these two
parameters are absent in the first column of the first order
approximation of $\delta M$, which describes the response of an
unpolarized source. Therefore, we suspect that the $Q$ and $U$ beam
dipole patterns may be due to leakage from the dipole pattern of $V$ into $Q/U$,
introduced by imperfect $\epsilon_{+}$ and $\varphi$ parameters we use in the boresight
calibration stage, as described in Section \ref{sec:pol_beam_pattern} 
  
%  between $V$ and $Q/U$, introduced by imperfect $\epsilon_{+}$ and $\varphi$ parameters.

We quantify the $Q$-$V$ and $U$-$V$
  linear correlations below to test this hypothesis.
We fit a linear regression model to the calibration spider scan data in the form of
$Q(x,y)+iU(x,y)=c+\tau V(x,y)+r(x,y)$, where $c$ and $\tau$ are a constant and the complex linear
coefficients, respectively, and $r$ the residual.  $\tau$ describes
the leakage fraction, $\tau=\delta M_{2,4}+i\delta M_{3,4}$ in
Eqn. (\ref{eq:deltaM}). The blue and green
lines in Figure \ref{fig:corr_QUV} show the real and imaginary part
of $\tau$, respectively.

\subsubsection{U-V Correction \label{sec:UV}}
\begin{figure}
\plotone{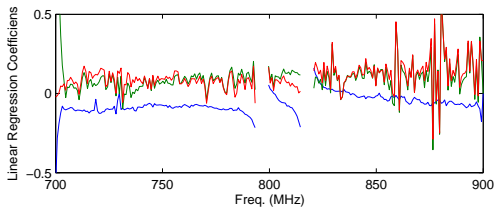}
\caption{$Q$-$V$ and $U$-$V$ correlations in spider scan data. The blue and green lines show the real and imaginary part of $\tau$. They also correspond to the $Q$-$V$ and $U$-$V$ correlations, respectively (see discussions in Section \ref{sec:QUV}). The red line indicates the pre-assumed $\arg(U+iV)$ of the noise diode, which is used to calibrate $\varphi$ for each night.  \label{fig:corr_QUV}}
\end{figure}

These relations could be intrinsic. They could also come from
inappropriate polarization calibration. The red line in Figure
\ref{fig:corr_QUV} indicates the assumed argument of $(U_\mathrm{ncal}+iV_\mathrm{ncal})$ derived from the noise diode Stokes parameters yielded by the analysis of the pulsar observations using the PSRCHIVE tools. There is a strong correlation between the slope of the linear $U$-$V$
relation (the imaginary part of $\tau$) and the PSRCHIVE-derived noise diode
argument.

As discussed in Section \ref{sec:pulsar}, we assume a noise diode
frequency profile $\vect{S}_\mathrm{ncal}(\nu )$ to solve for $G$, $\gamma$, and
$\varphi$ for each observing session. 
Looking at the third and fourth rows of Eqn. (\ref{eq:M0_ncal}), it can
be shown that $\varphi =0.5 \times \arg[(U^{'}_\mathrm{ncal}+iV^{'}_\mathrm{ncal})/(U_\mathrm{ncal}+iV_\mathrm{ncal})]$.
Combined with Eqn. (\ref{eq:deltaM}), one can see that a biased $\delta\arg(U_\mathrm{ncal}+iV_\mathrm{ncal})$ will lead to
a $\delta U=V\times\delta \arg(U_\mathrm{ncal}+iV_\mathrm{ncal})$.
Therefore, the strong correlation between $\arg(U_\mathrm{ncal}+iV_\mathrm{ncal})$
and the imaginary part of $\tau$ implies that $\delta
\arg(U_\mathrm{ncal}+iV_\mathrm{ncal})$ is responsible for the correlation between
$U$ and $V$. In fact, after setting $\arg(U_\mathrm{ncal}+iV_\mathrm{ncal})=0$, the
correlation between the $U$ and $V$ beam disappears.

\subsubsection{Q-V Correction\label{sec:Pol_Beam_corr}}

On the other hand, the $Q$-$V$ relation could be also from errors in the derived Mueller matrix
parameters. Assuming errors in $\epsilon_{+}$ is the only source of
the observed $Q$-$V$ coupling in spider scan data,  we estimate
$\delta\epsilon_{+}$ from 3C295 spider scan data, and apply the
`corrected' $\epsilon_{+}$ to spider scan data of other unpolarized sources, including 3C147 and 3C295.
It appears that the dipole features in $Q$ are removed by the correction of $\epsilon_{+}$. 

However, the removal of dipole shapes in $Q$ does not answer the question: Does the dipole shape come from 
leakage of $V$ to $Q$ due to $\delta\epsilon_{+}$? Or is it an intrinsic beam feature?
To distinguish these two possible scenarios, we apply the modified $\epsilon_{+}$ to the polarized 3C286 data.

The linear polarization fraction of 3C286 is $\sim 5\%$, and the circular polarization is negligible.
One can expect that $\delta\epsilon_{+}$ mainly influences $V$ in the calibrated 3C286 data. 
Comparing the RMS of $V$ over tracking scans with different parallactic 
and Faraday rotation angles, we may have a handle on the origin of the
$Q$-$V$ coupling.

We use 57 scans of 3C286 taken in 2011 to perform the test. The angles
of rotation $\phi$ caused by sky rotation and ionospheric Faraday
rotation range between $-88.4^{\circ}$ to $60.6^{\circ}$. 
We calculate $\sigma_{V}(\nu)$, the RMS of calibrated $V$ over 57
tracking scans at each frequency channel,  discard the highest $10\%$
$\sigma_{V}(\nu)$ values to minimize impact of RFI and calculate the
mean values.  With $\epsilon_{+}$ derived from pulsars and subsequently modified by
spider scan data, the mean value of $\sigma_{V}$ reduces from $0.752\% \pm 0.016\%$ to $0.716\% \pm 0.018\%$.
The improvement appears marginal. 

\begin{figure*}
\plotone{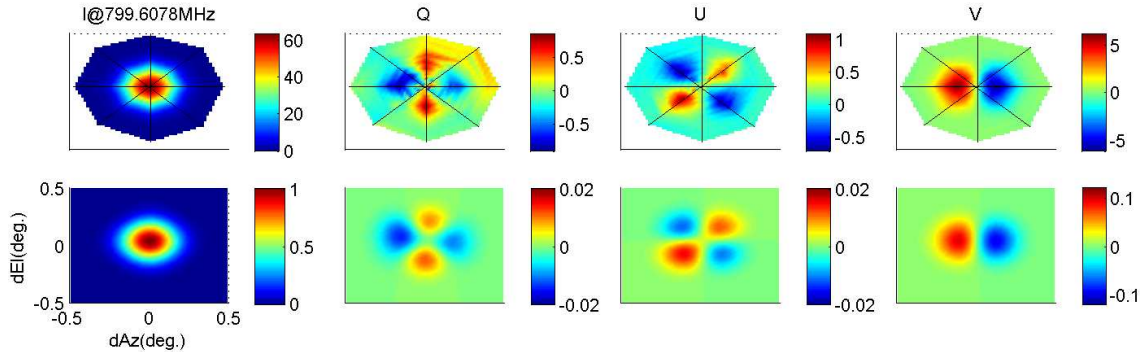}
\caption{Polarized beam pattern at 800 MHz derived from spider scans of
  3C295. The patterns in the upper panels are linear interpolation over the calibrated data of the spider scan, and
  the lower ones show the best fit results of the Gauss-Hermite
  model. The color scales of the upper panels are temperature in units
  of Kelvin, while the lower panels show fractional intensity compared
  to central source 3C295.  Note we use modified Mueller matrix
  parameters as described in Section \ref{sec:QUV} and
  \ref{sec:Pol_Beam_corr}. As a result, the dipole patterns in the $Q$
  and $U$ beams, first reported in Section \ref{sec:QUV}, have been mitigated in this figure. \label{fig:spider_scan_map}}
\end{figure*}

\begin{figure}
\plotone{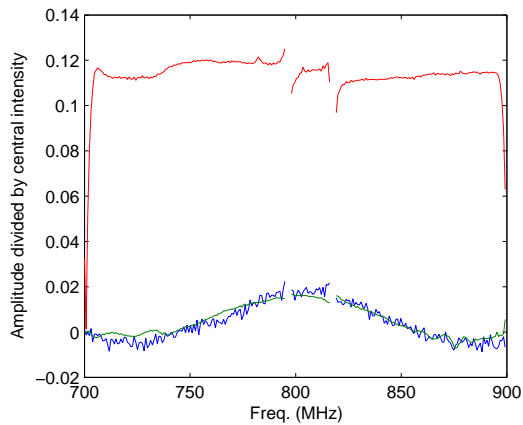}
\caption{Coefficients of spider scan beam maps including quadrupole term of $Q$(blue), quadrupole term of $U$(green) and dipole term of $V$(red). The coefficients here are normalized by the peak value of $I$ beam, and averaged over 7 spider scans on 3C48, 3C147, and 3C295. \label{fig:Quadrupole}}
\end{figure}

Figure \ref{fig:spider_scan_map} shows examples of the polarization
beam pattern obtained from quasar spider scans, after correcting the
$\varphi$ and $\epsilon_{+}$ parameters. In addition to the dipole pattern
in the $V$ beam, the $Q$ and $U$ beams show a less prominent
quadrupole feature in the frequency range of 750-850 MHz. The
amplitudes of these
features are plotted as a function of frequency in Figure \ref{fig:Quadrupole}.

\subsection{Polarization Beam from Pulsar tracking scans}\label{sec:pulsar_beam}
Unpolarized quasar data provide useful information on the first column
of $\matr{M}(\Delta \vec{x})$, and polarized sources are required in order to characterize the entire
Mueller matrix.  However, polarized quasars, such as 3C286 with a
polarization fraction of about $5\%$, cannot provide significant
signatures. For example, the expected leakage beam pattern from $Q$ to
$I$, according to the measured Mueller matrix model, is about $2\%$. With parallactic angle rotation,
the change in $Q$ of a $5\%$ linearly polarized source is no greater than $10\%$ of total intensity.
The signature of $Q$ leakage to $I$ is therefore $0.2\%$ at most, below our calibration
significance of $0.6\%$. Furthermore, variations of the diffuse
Galactic foreground radiation in the primary beam can contaminate the
observation;   it is difficult to completely separate quasar signals 
from the diffuse foreground, given a primary beam FWHM of 15' at 800
MHz at the GBT. We therefore solve for the full Mueller matrix parameters
of the primary beam with off-source pulsar data;  by subtracting
off-pulse phase bins from the on-pulse data, we mitigate the diffuse
background while preserving the pulsar signals.

Assuming a known pulsar profile, we solve for the full Mueller matrix
with the primary beam pattern $\matr{M}(\Delta\vec{x})$ at discrete positions,
$\vec{x_{0}}$, assuming $\vect{S}(\vec{x'})$ is a delta function peaking at $\vec{x_{0}}$ in
Eqn. (\ref{eq:s'sp}).

We take 59 tracking scans of pulsar B1133+16, including 27 on-source tracking scans,
and 32 off-centered ones with positions spread within a $0.19^\circ$
radius from the beam center. The radius of $0.19^\circ$ is about $0.7$
times the beam FWHM at 800 MHz.  The B1133+16 pulsar profile is
obtained from multiple parallactic angle observations as described in
Section \ref{sec:pulsar} and use PSRCHIVE matrix template matching mode described in \cite{Straten2006} and \cite{Straten2013}
to solve for the
Mueller matrix parameters for each on- and off-source tracking scan. We allow
$\theta_{0}$ to vary to account for possible relative $Q$-$U$ rotation
between beam center and other parts of the beam.

After solving the Mueller parameters at 33 locations (one on-source
and 32 off-centered ones), we again perform a 2D Gauss-Hermite fit 
(Eqn. (\ref{eq:gaussian_hermite_function})) to model the beam. Due
to observational noise and variations of the pulsar profile itself,
we are not able to model at high confidence level the beam pattern for
individual channels. Instead, we average the Mueller parameters over
the frequency range 755-845 MHz for the Gauss-Hermite fitting. 

\begin{figure*}
\plotone{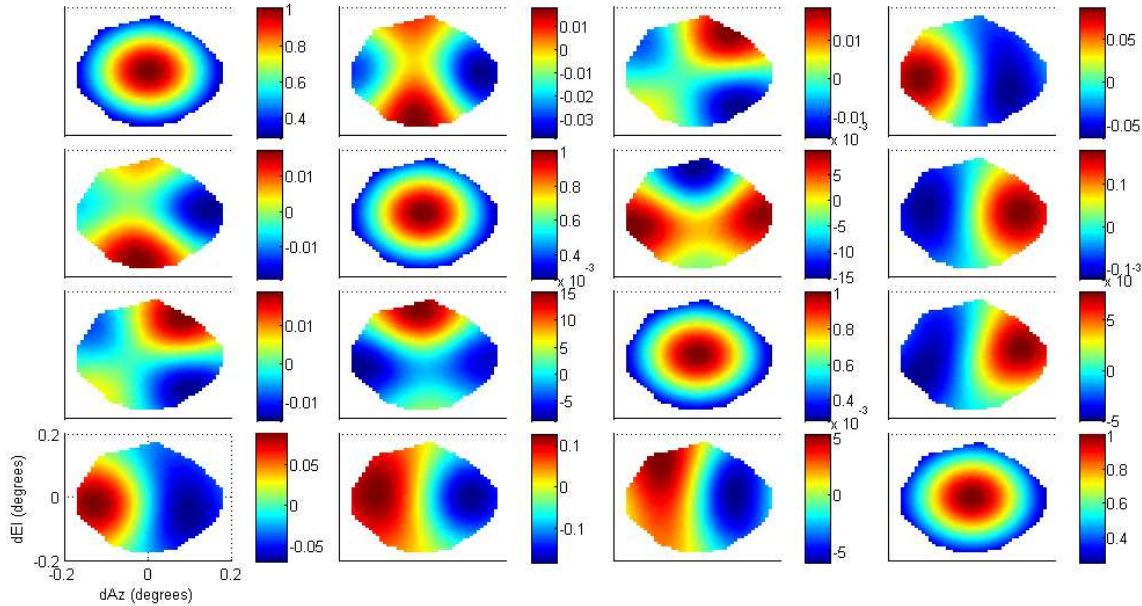}
\caption{Beam pattern of the $4 \times 4$ Mueller matrix elements
  fitted with off-centered pulsar tracking scans. The diagonal elements are
  normalized to one, and from
  top-left to bottom-right the diagonal panels correspond to $M_{II}, M_{QQ},
  M_{UU}, M_{VV}$, respectively. \label{fig:pulsar_beam_m}}
\end{figure*}

Figure \ref{fig:pulsar_beam_m} shows the Mueller matrix beam calculated with 
best fit results of parameters. As 3C295 is unpolarized, the first column of Figure \ref{fig:pulsar_beam_m}
correspond to spider scan results on 3C295. Comparing Figure
\ref{fig:spider_scan_map} and Figure \ref{fig:pulsar_beam_m}, we find
similar beam patterns with leakage derived from pulsar data and 
quasar spider scans.

To verify the goodness of fit, we perform a `significance test' on each Jones
matrix parameter. We simulate 33 sets of Mueller parameters, each
with a normal distribution and a zero mean, and randomly assign them
to the 33 locations. We again perform the 2D Gauss-Hermite fits, and
record $R^2\equiv 1-(\sigma_\mathrm{residual}/\sigma_\mathrm{data})^2$, where
$\sigma_\mathrm{data}$ and $\sigma_\mathrm{residual}$ are the standard deviations of the
33 simulated input and fitted parameters, respectively.  With 500 such simulations, we derive the probability
$P(R^2)$ that random parameters get better fitting results. It appears only
the $\epsilon_{-}$ fit, which describes the dominant Stokes $I$ to $V$ leakage, is significant ($P<5\%$). Although $\gamma$ and
$\theta_{-}$ show similar patterns as $Q$ and $U$, respectively, they do not pass the significance test. 

According to the spider scans, the expected patterns of Stokes $I$ leakage into $Q$ and $U$ are less than one
sixth in amplitude of Stokes $I$ leakage into $V$, and are comparable to the
uncertainty of polarization fraction of pulsar B1133+16, as shown in
Figure \ref{fig:RMS_p_f_B1133}. Therefore, it is not entirely surprising
that we cannot detect significant patterns of $\gamma$ and
$\theta_{-}$.  Pulsar B0450+55 also shows a similar level of fluctuations
in polarization fraction, unfortunately. 

\section{Procedure of On-axis Polarization Calibration}\label{sec:summary}
The essential task for on-axis polarization calibration in this work is determining the parameters $G$, $\gamma$, $\varphi$, $\theta_{-}$, $\epsilon_{+}$ and $\epsilon_{-}$ as functions of time, and $\phi_\mathrm{RM}$ as function of time, azimuth, and elevation. Here we will summarize the methodology we adopt to determine these parameters. \\ \\
1. Estimate noise diode frequency profile for each session with data of on/off source tracking scans on unpolarized quasars, like 3C48, 3C147, and 3C295 (Section \ref{sec:ncal_profile}). \\ \\
2. Estimate $G$ and $\gamma$ by comparing observed noise diode data with estimated noise diode profile (Section \ref{sec:G_gamma_varphi}). \\ \\
3. Estimate $\varphi$ from observed $\arg(U^{'}_\mathrm{ncal}+iV^{'}_\mathrm{ncal})$ of noise diode with assumed $V_\mathrm{ncal}=0$ in real noise diode signal. (Section \ref{sec:UV}). \\ \\
4. Apply $G$, $\gamma$, and $\varphi$ to calibrate unpolarized quasar data (with the other parameters set to be $0$). Then modify $\gamma$, $\theta_{-}$, and $\epsilon_{-}$ based on quasar data calibrated in this step (Section \ref{sec:1st_order}). \\ \\
5. Apply $G$, $\varphi$, along with modified $\gamma$, $\theta_{-}$, and $\epsilon_{-}$ to calibrate spider scan data on another unpolarized quasar. Then estimate $\epsilon_{+}$ with correlation between dipole features in $Q$ and $V$ data (Section \ref{sec:Pol_Beam_corr}). \\ \\
6. Estimate $\phi_\mathrm{RM}$ with EOFs, US-TEC data, and IGRF. Then do ionospheric RM correction accordingly (Section \ref{sec:RM}).

\section{Discussion and Conclusion}\label{sec:conclusion}
In this paper, we present the polarization calibration methodology for
the GBT HI intensity mapping experiment.  Accurate polarization
calibration is critical to properly mitigate the unwanted leakage from polarized synchrotron foregrounds
into total intensity. 

We use multiple parallactic angle pulsar observations to solve for
the six Jones Matrix parameters at boresight in each of the 256 frequency channels
between 700-900 MHz at the GBT.  Applying the solutions to unpolarized quasar
observations, the RMS fluctuations of the Stokes
parameters over time are about $1.3\%$-$1.7\%$ of total intensity.  As a first-order correction, some of
the Jones Matrix parameters are further modified based on tracking and spider scans of
quasars. The RMS fluctuation reduces to $0.6\%$-$0.8\%$ of total
intensity after the correction.

In Section \ref{sec:pulsar} we discuss that there are
two ways to break the degeneracy between $I$ and $V$ in PSRCHIVE:  One assumes
$V$ to be negligible while observing a standard calibrator, i.e., the
sum of $V$ from the astrophysical calibrator, the (diffuse) sky, and
the system, is negligible.   However, the $\epsilon_{-}$ parameter
obtained from PSRCHIVE under this assumption, which mainly describes the leakage between
$I$ and $V$, appears to be incorrect. This assumption does not seem to be valid in our case.

The other approach is to assume the $V/U$ ratio of the noise diode is
known.  Section \ref{sec:QUV} and Figure \ref{fig:corr_QUV}
suggest that this could be a good assumption.  However,  PSRCHIVE
assumes the noise diode signal is injected early in the system so
that it shares the same light path as the sky signal;  this is 
not the case for our GBT 800MHz observations, since the noise diode
signal is injected after the OMT as discussed in Section \ref{sec:instrument}.

Another explanation for the imperfection might be the variation of polarized pulsar profiles. PSR
B1133+16, one of the pulsars we use to solve the Jones Matrix
parameters, has been reported to have 
``orthogonally polarized
modes'', which may be responsible for the variations in the integrated
fractional polarization and position angle of the pulsar
\citep{Karastergiou2002}. The polarized pulse or frequency profiles of pulsars can
in general fluctuate due to the astrophysical complexity of the pulse
mechanism, or interference of the interstellar medium. 

The polarized frequency profiles of quasars are more
stable. However, with high fractional linear polarization and
non-negligible circular polarization, pulsars provide more information
needed to solve for all the Jones matrix model parameters than
quasars, which are usually slightly or not polarized. Some of the
parameters, including $\varphi$, $\epsilon_{+}$, and $\theta_{+}$,
cannot be constrained by unpolarized sources. We can determine
$\theta_{+}$ by comparing the calibrated polarization position angle
of 3C286 with the known value, while $\varphi$ can be constrained by a
highly polarized noise diode, which is however found to contain
an uncertain $V/U$ spectrum.  The $\epsilon$ parameter is not affected by the noise
diode as the signal injection takes place after the dipole receptor in
the signal stream at the GBT.  

Off-centered polarization leakage is also an important source of
contamination.  With quasar spider scans, we find a dominant dipole
feature in the Stokes $I$ to $V$ leakage pattern, which is at the $\approx
12\%$ level, and secondary quadrupole features of Stokes $I$ to $Q$ and I
to $U$ leakage patterns, which are $\lesssim 2\%$ of total
intensity. With the leakage of $V$ dipole feature into $Q$ and $U$, we can
estimate $\delta\varphi$ and $\delta\epsilon_{+}$, and improve the calibration of polarized sources, like 3C286. Although there are still potential intrinsic dipole features in $Q$ and $U$ beam pattern which have yet to be separated from leakage of $U$.

We find similar features using off-centered pulsar observations and map
out the entire Mueller matrix primary beam. However, the Stokes $I$ to $V$
leakage appears to be the only significantly determined beam pattern. The beam features
of $I$ leakage to $Q$ and $U$ are comparable to the variation of
polarization profile of PSR B1133+16. Therefore, it is not surprising
that we cannot significantly measure these features in the pulsar data.

In this paper, we measured the RMS fluctuations of calibrated on-source data of
quasars, including unpolarized 3C48, 3C295, and 3C147, and slightly
polarized 3C286, to be $0.6$-$0.8\%$ of the total intensity. We also
mapped the polarization beam pattern. Accurate polarization calibration at this
level is required to mitigate the polarized foreground contribution
for HI intensity mapping power spectrum measurements.  We will report
improvements on the redshifted HI power spectrum in future work. We will also investigate the
Faraday rotation measure (RM) synthesis of Galactic foregrounds in the HI intensity mapping fields.

\acknowledgments
We thank Willem van Straten and Paul Demorest for their invaluable help with the
use of PSRCHIVE and pulsar calibration with the GUPPI backend at the
GBT.  We thank the anonymous referee for the thorough comments
that improved the clarify of the manuscript. T.-C. C. acknowledges support from MoST grant 103-2112-M-001-002-MY3.

\appendix
\section{First-order approximation of the polarization calibration error}
In the model we use, $\matr{J}$ is a function of seven Jones matrix
parameters, which can be rewritten as $\matr{J}(\vec{p})$, where
$\vec{p}=\{p_{i}\}=\{G, \gamma, \varphi, \theta_{+}, \theta_{-},
\epsilon_{+}, \epsilon_{-}\}$ is the parameter vector. $\epsilon_{+}$
and $\epsilon_{-}$ are defined as
$\epsilon_{+}=\epsilon_{0}+\epsilon_{1}$ and
$\epsilon_{-}=\epsilon_{0}-\epsilon_{1}$.  Mueller matrix can also be
written as a function of these parameters $\matr{M}(\vec{p})$.  See
Section \ref{sec:model} for details.

Ideally, the polarization calibration should recover the real signal
by applying inverse matrices of $\matr{M}$ and $\matr{R}(\phi)$ to Eqn.
(\ref{eq:s'sp}). However, if the estimated parameter vector
$\vec{p^{'}}=\vec{p}+\delta\vec{p}$ is slightly different from real
parameter vector $\vec{p}$, the calibration procedure becomes 
\begin{eqnarray}
 \vect{S}^{"}_\mathrm{sp}&=&\matr{R}^{-1}\matr{M}(\vec{p^{'}})^{-1}\vect{S}^{'}_\mathrm{sp} \nonumber \\
                  &=&\matr{R}^{-1}\matr{M}(\vec{p^{'}})^{-1}\matr{M}(\vec{p})\matr{R}\vect{S}_\mathrm{sp} \nonumber \\
                  &=&\vect{S}_\mathrm{sp}+\matr{R}^{-1}\delta\matr{MR}\vect{S}_\mathrm{sp},
                  \label{eq:S''}
\end{eqnarray}
where $\delta \matr{M}=\matr{M}(\vec{p^{'}})^{-1}\matr{M}(\vec{p})-\matr{I}$, $\matr{I}$ is the identity
matrix. The term $\matr{R}^{-1}\delta \matr{MR}\vect{S}_\mathrm{sp}$ is then the error introduced
by the polarization calibration procedure.

If $\theta_{+}, \theta_{-}, \epsilon_{0}, \epsilon_{1}\ll1$, the first-order perturbation of $\matr{M}$ is 
\begin{eqnarray}
\delta \matr{M}\approx\left(\begin{array}{cccc}
-2\delta G        &-2\delta\gamma     &-\delta\theta_{-} &\delta\epsilon_{-}\\       
-2\delta\gamma    &-2\delta G         &0                &\delta\epsilon_{+}\\
-\delta\theta_{-}  &0                  &-2\delta G       &-2\delta\varphi   \\
\delta\epsilon_{-}&-\delta\epsilon_{+}&2\delta\varphi   &-2\delta G
\end{array}\right).
\label{eq:deltaM}
\end{eqnarray}
We can then write down the expression for the error,  
\begin{eqnarray}
\begin{array}{ll}
\matr{R}^{-1}(\phi+\delta\phi)\delta \matr{MR}(\phi)\approx &   \\   
 
\\ \left(\begin{array}{cccc}
-2\delta G                                      &-(2\delta\gamma\cos2\phi-\delta\theta_{-}\sin2\phi)                                &-\delta\theta_{-}\cos2\phi-2\delta\gamma\sin2\phi  &\delta\epsilon_{-}\\       
-(2\delta\gamma\cos2\phi-\delta\theta_{-}\sin2\phi) &-2\delta G                                                                            &-2\delta\phi                                       &\delta\epsilon_{+}\cos2\phi+2\delta\varphi\sin2\phi\\
-\delta\theta_{-}\cos2\phi-2\delta\gamma\sin2\phi   &2\delta\phi                                                                                              &-2\delta G                                        &\delta\epsilon_{+}\sin2\phi-2\delta\varphi\cos2\phi   \\
\delta\epsilon_{-}                                 &-(\delta\epsilon_{+}\cos2\phi+2\delta\varphi\sin2\phi)&                      -\delta\epsilon_{+}\sin2\phi+2\delta\varphi\cos2\phi &-2\delta G
\end{array}\right).
\end{array}
\label{eq:RdeltaMR}
\end{eqnarray}
The operator $\matr{R}$ can be combined with the effect of $\theta_{+}$ (see section \ref{sec:model}), so $\delta\theta_{+}$ behaves like $\delta\phi$ above.

\end{document}